\documentclass[12pt,english,review]{elsarticle}
\usepackage[T1]{fontenc}
\usepackage[latin9]{inputenc}
\usepackage{geometry}
\usepackage{amsmath}
\usepackage{graphicx}
\usepackage{setspace}
\usepackage{esint}
\makeatletter

\providecommand{\tabularnewline}{\\}

\usepackage{array}
\usepackage{multirow}

\usepackage{babel}

\usepackage{babel}

\usepackage{babel}

\@ifundefined{showcaptionsetup}{}{%
 \PassOptionsToPackage{caption=false}{subfig}}
\usepackage{subfig}
\makeatother

\usepackage{babel}
\begin{document}
\begin{frontmatter}

\title{Experimental study of forces on freely moving spherical particles
during resuspension into turbulent flow}

\author{Hadar Traugott}

\author{Alex Liberzon\corref{cor1}}

\ead{alexlib@eng.tau.ac.il}

\address{School of Mechanical Engineering, Tel Aviv University, Tel Aviv 69978,
Israel }

\cortext[cor1]{Corresponding author}
\begin{abstract}
Turbulent resuspension, a process of lifting solid particles from
the bottom by turbulent flow, is ubiquitous in environmental and industrial
applications. The process is a sequence of events that start with
an incipient motion of the particle being dislodged from its place,
continue as sliding or rolling on the surface, ending with the particle
being detached from the surface and lifted up into the flow. In this
study the focus is on the resuspension of solid spherical particles
with the density comparable to that of the fluid and the diameter
comparable with the Kolmogorov length scale. We track their motion
during the lift-off events in an oscillating grid turbulent flow.
We measure simultaneously the Lagrangian trajectories of both the
particles freely moving along the bottom smooth wall and the surrounding
flow tracers. Different force terms acting on particles were estimated
based on particle motion and local flow parameters. The results show
that: \emph{i}) the lift force is dominant; \emph{ii}) drag force
on freely moving particles is less relevant in this type of resuspension;
\emph{iii}) the Basset (history or viscous-unsteady) force is a non-negligible
component and plays an important role before the lift-off event. Although
we cannot estimate very accurately the magnitude of the force terms,
we find that during the resuspension they are within the range of
$2\div10$ times the buoyancy force magnitude. The findings cannot
be extrapolated to particles, which are much smaller than the Kolmogorov
length scale, or much denser than the fluid. Nevertheless, the present
findings can assist in modeling of the sediment transport, particle
filtration, pneumatic conveying and mixing in bio-reactors. \end{abstract}
\end{frontmatter}

\section{Introduction}

Resuspension is the process of particle release from a surface into
a surrounding fluid flow. In order to distinguish the entrainment
into a flow from a motion along the surface (sliding or rolling),
it is often denoted as ``lift-off'' \citep{Brennen:2005,Rijn:1984,henry_minier:2014}.
Resuspension of particles is an important mechanism in a large variety
of practical applications such as particle filtration \citep{Huang:2008},
oil production \citep{middletich:1981}, contamination in clean rooms
\citep{dixon:2006}, pneumatic conveying \citep{soepyan:2016} and
particle behavior in respiratory ways \citep{Sarma:1992}. In order
to predict particle resuspension accurately, the relation of the incipient
motion and the removal of particles from surfaces to the particle/fluid
properties and the local flow regime, need to be understood in detail.
Prediction of lift-off events in turbulent flows requires understanding
of both the surrounding flow field and the particle-flow interaction.
The latter is based on the balance of forces and moments resulting
from stress applied on the particle by the local flow and the restrictive
forces of gravity and surface/particle interactions\citep{Ziskind:2006,henry_minier:2014}.

Despite numerous experimental and numerical studies addressing the
problem of incipient motion in general, and lift-off in particular,
the question on which mechanism dominates the process remains open.
Several studies propose that particle motion is predominantly driven
by the magnitude of fluctuating drag and lift forces exerted on particles,
depending on their degree of exposure to the flow \citet{Schmeeckle:2007,Dwivedi:2011}.
The importance of instantaneous fluctuating velocities might indicate
that the turbulence structure at the near bed is ultimately responsible
for particle motion \citep{McLean:1994}. \citet{Gimenez-Curto:2009}
suggested that the critical motion is related to the maximum forces
acting on the particles, rather than the mean bed shear stress. \citet{celik:2010}
found that the time duration of the force above a certain threshold
is the critical parameter.

In respect to the direction of the force acting on particles fixed
on the bottom, the discussion is divided between the drag force (a
component parallel to the streamwise flow velocity) and the lift (perpendicular,
vertical) force. For turbulent flows with the streamwise flow direction
parallel to the bottom bed surface, the hydrodynamic drag term is
considered to be dominant. For instance, \citet{Schmeeckle:2007}
used a force transducer directly connected to a particle to measure
force synchronously with the flow velocity measurements above or in
front of the particle. The horizontal force was shown to correlate
with the magnitude of the downstream velocity but not with the vertical
velocity component. In this experiment, the standard drag model based
on the streamwise velocity predicted the horizontal force acting on
a fixed and fully exposed particle \citep{Schmeeckle:2007}. The vertical
force, on the contrary, correlated poorly with both horizontal and
vertical velocity components. Similarly, \citet{Nelson:1995} have
reported strong correlation between sediment rate (number of resuspended
particles) and streamwise velocity fluctuations near the bed, opposite
to weak correlation with the fluctuations of vertical velocity. \citet{Molinger_Nieuwstadt:1996}
studied the lift force and could confirm experimentally the predictions
of the mean lift force of \citet{Hall:1988}. In a similar type of
study, \citet{Dwivedi:2011} used a force sensor attached to a fixed
particle. The authors concluded that the lift force is produced primarily
by the pressure gradient in the flow due to externally imposed unsteadiness
of the flow or turbulent fluctuations. \citet{Dwivedi:2011} suggested
a particular local turbulent flow structure that could produce high
pressure below the particle and low pressure on above it, leading
to high lift force. A similar mechanism was proposed by \citet{Zanke:2003}
as a possible cause of particle suspension. 

It is noteworthy that the aforementioned studies measured forces on
fixed particles. Moreover, due to the resolution of force sensors
the particles were relatively large. There is no information on the
forces applied by a turbulent flow on freely moving particles. Recently
we have developed the necessary tools for such measurements. In \citet{Shnapp:2015}
the trajectories of spherical particles lifted off smooth and rough
surfaces were measured in a tornado-like vortex flow. \citet{meller:2015}
have extended the \citet{Sridhar_Katz:1995} method to measure relative
velocity between a particle and the surrounding turbulent flow and
estimate various components of force acting on suspended particles.
Combining these two developments with the oscillating grid apparatus
presented by \citet{Traugott:2011}, in this study we estimate the
force components acting on particles freely moving on the bottom wall
and the relative contributions of force components to the lift-off
events.

We utilize the three-dimensional particle tracking velocimetry (3D-PTV)
to obtain velocity and acceleration data along trajectories of tracers
and large spherical particles, as described in Section$\,$\eqref{sec:Materials-and-methods}.
The method provides parameters of individual, freely moving solid
particles before, during, and after the lift-off events. These measured
simultaneously with the local turbulent flow, represented by Lagrangian
tracer trajectories. We focus on the moment of lift-off of the particle
from the wall and its relation to the local turbulent flow characteristics.
Applying the particle equation of motion we estimate the magnitude
and direction of inertia, pressure, drag, lift and Basset force terms
and their effect on the lift-off events. Experimental results are
presented in Section~\eqref{sec:Results-and-discussion} and discussed
in the closing Section~\eqref{sec:Summary-and-conclusions}.

\section{Methods and materials \label{sec:Materials-and-methods}}

\subsection{Experimental setup\label{sub:Experimental-setup}}

\begin{figure}
\begin{centering}
\includegraphics[width=0.6\columnwidth]{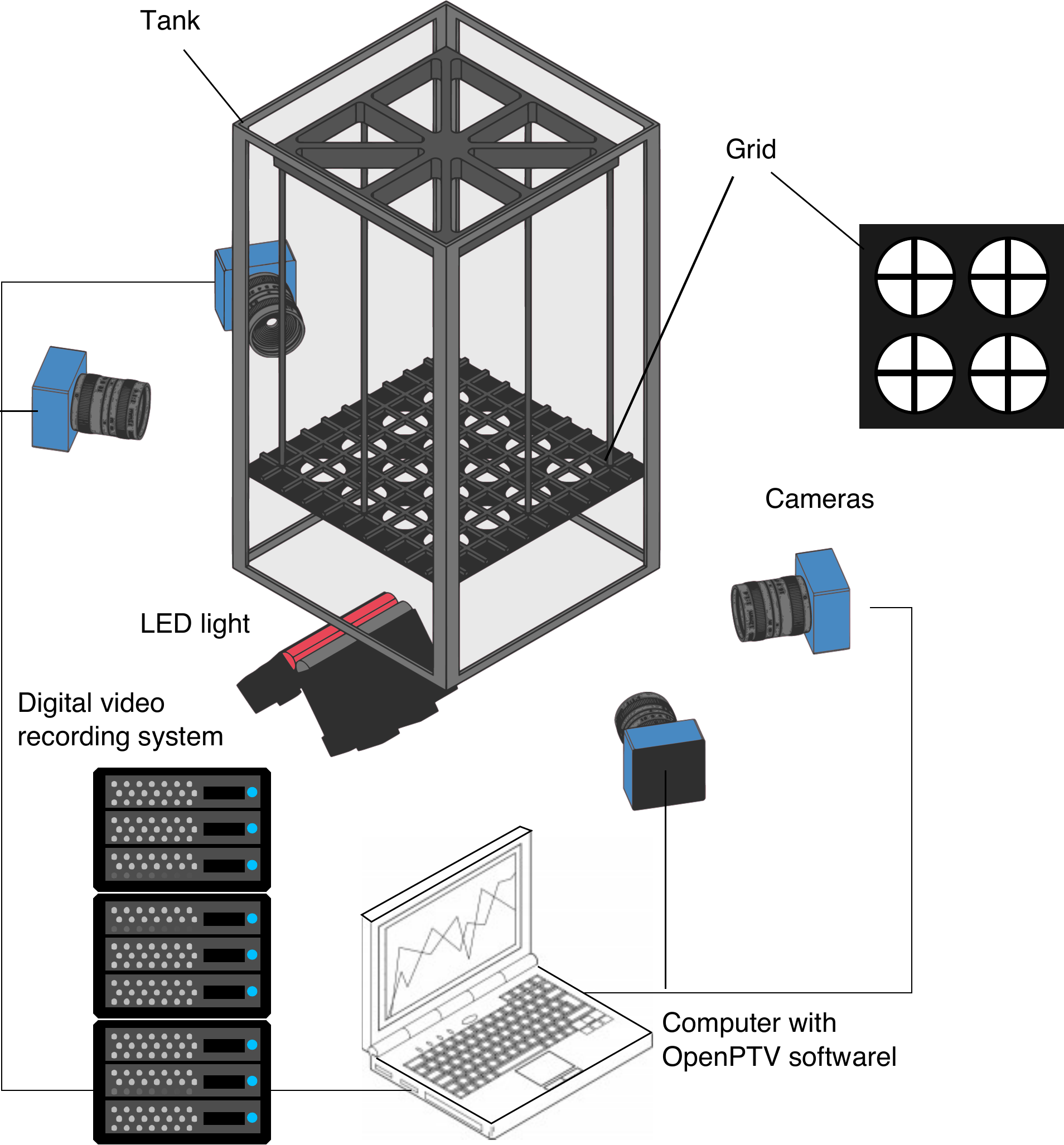} 
\par\end{centering}

\protect\protect\caption{Schematic view of vertically oscillating grid in a glass tank. The
two LED light sources provide a volumetric illumination of the flow
field under the grid. Four high-speed CMOS cameras synchronously record
the motion of the particles and flow tracers on a real-time video,
streamed to a digital video recording system (IO Industries Inc.).
\label{fig:exp_setup}}
\end{figure}

The oscillating grid setup is shown schematically in Fig.~\eqref{fig:exp_setup}.
The system comprises of a glass tank ($30\times30$ cm and $50$ cm
tall) and a vertically oscillating grid on an eccentric shaft driven
by a $1.5\ \mathrm{kW}$ variable speed electrical motor (CDF90L-4,
KAIJIELI Inc.). The tank was filled with filtered tap water until
a height of $220$ mm, the grid height was set within the range of
$h=100\div101$ mm (measured from the bottom of the chamber) and stroke
amplitude (peak-to-peak) $s_{l}=10$ mm. The frequency of oscillation
of the grid is controlled by changing the input voltage to the motor.
We present here the results of the runs at 1.5, 1.7, 1.2 and 2.1 Hz
. The grid, shown in a top view in Fig.~\eqref{fig:exp_setup} was
made of square bars covered by a plastic sheet with $4\times4$ arrangement
of circular holes, in order to increase the grid solidity to 80\%.
Lower solidity grids did not create a sufficient number of particle
lift-off events. 

Prior to the 3D-PTV study, the flow field under the grid was characterized
using particle image velocimetry (PIV), see \citet{Traugott:2011}.
The PIV data allows to define the important scales of turbulent flow
under the oscillating grid. The Kolmogorov length scales varied within
the range $\eta=(\nu^{3}/\varepsilon)^{1/4}=270\div330\:\mu\textrm{m}$
and the Kolmogorov time scales estimated as $\tau_{\eta}=\left(\eta/\nu\right)^{1/2}=0.07\div0.1$
seconds. The flow under the grid was neither homogeneous nor isotropic
as it might be expected due to the relative proximity of the grid
to the bottom wall and the large solidity. 

The 3D-PTV experimental system shown schematically in Fig.~\eqref{fig:exp_setup}
included the following components: the digital video recording system
(IO Industries Inc.) and four high speed CMOS cameras ($8\ \mathrm{bit}$,
$1280\times1024\ \mathrm{pixels}$, EoSens GE, Mikrotron), equipped
with $60$ mm lenses (F-mount, Nikon). The cameras capture simultaneously
(with a maximum possible time jitter of $0.001\ \mathrm{fps}$) digital
video recorded at the rate up to $700\ \mathrm{M\mathrm{b/s}}$ on
high-speed hard drives. The data is analyzed using the open source
software, ``OpenPTV'' \citep{openptv}. 

Cameras were located in an angular array from two sides of the grid
chamber, as shown in Fig.~\eqref{fig:exp_setup}. The four cameras
arrangement reduces the number of ambiguities and allows reliable
determination of most of the particles which are completely hidden
in one of four images \citep{Dracos:1996}. In the particular runs
reported here, image acquisition rate was set to $160$ fps, the observation
volume was $7.5\times4\times6$ cm$^{3}$ (length $\times$ width
$\times$ height). Two light emitting diodes (LED) line sources (Metaphase,
USA) illuminated the observation volume in the center of the grid
chamber. The combination of the two LED light sources provided a nearly
uniform light intensity across a wider observation volume. A two step
calibration method was used: a static calibration using a three-dimensional
reference target, and a dynamic calibration using a dumbbell (wand)
moving in a measurement volume after the grid was installed in the
operating condition~\citep{openptv}.

Two different types of particles were used in order to obtain simultaneous
recordings of the turbulent flow and the motion of the inertial particles
in the same observation volume. Polyamide particles with a mean diameter
of $50\ \mathrm{\mu m}$ and density of $1030$ kg/m$^{3}$ (Dantec
Dynamics Inc.) were used as tracers. The relaxation time of flow tracers,
$\tau_{p}=\rho_{p}d_{p}^{2}/18\mu$ is approximately $0.143$ ms,
which is significantly smaller than the Kolmogorov time scale of the
flow and their Stokes number is small, $St<0.01$. Furthermore, the
particles fulfill the conservative restrictions for flows with acceleration
of less than $10\ \mathrm{m/s^{2}}$ e.g. \citet{Dracos:1996} (see
class of neutrally buoyant, $d_{p}<60\ \mathrm{\mu m}$ particles),
and therefore behave as tracers in a given turbulent flow. 

Silica gel spheres (Fulka Inc.), $550$ $\mu m$ in diameter (of the
order of magnitude of the Kolmogorov length scale) and effective density
of $\rho_{p}=1062$ kg/m$^{3}$ were used as the inertial particles.
Their Stokes number is approximately equal to 0.2. 

Before each experimental run, approximately 30 inertial particles
were spread randomly on the bottom. Each run started in stagnant water,
and after a certain time the inertial particles were entrained into
the water column by the turbulent flow generated under the constantly
oscillating grid. For the chosen frequencies ($1.5,1.7,1.9$ and $2.1$
Hz), all inertial particles became suspended. The number of particles
spread on the bottom of the tank is relatively low and they are lifted-off
separately, without particle-particle interactions. 3D-PTV analysis,
including particle identification and tracking, was performed on the
inertial particles separately from the tracers using the same video
files. The software \citep{openptv} allows to discriminate between
the particles and the tracers using their size.\emph{ }

It is important to note that the number of ``successful| events of
inertial particles that participate in the statistics in different
runs were only 8 $\div$ 20. These are neither the total number of
the events nor the total number of trajectories recorded by 3D-PTV.
The 30 particles spread on the bottom wall of a tank move freely in
all possible directions. Some particles reach the corners of the tank,
while a majority is freely moving in the tank going through numerous
suspension - deposition cycles. Nevertheless, the number of ``successful''
events relate to a very small sample selected by post-processing.
These events: a) occur within the field of view of the 3D-PTV system
(many particles are lifted outside of the volume or start their motion
inside and lift-off ends outside of the volume, etc.), b) are successfully
tracked during the full length of the event (due to the restrictions
of Basset force estimate we had to have the particle within the field
of view for at least 10 time steps prior to upward motion), c) the
particles are lifted-off and not deposited for at least 10 Kolmogorov
times (otherwise the event is denoted as saltation) , d) have sufficient
tracer density surrounding the particle before, during and after the
lift-off event. Due to this large set of complex restrictions, the
amount of ``successful'' events that we can add into the statistics
is relatively small. Collection of more data was impossible due to
the large time span of the experiment and data analysis.

\subsection{Particle equation of motion\label{sub:Particle's-equation-of}}

The method used herein to estimate the forces acting on the particles
is described in details in \citet{meller:2015}, and briefly reproduced
here for completeness. The main difference is the estimate of the
Basset force as explained below. The method is based on the equation
of motion of a point-like sphere in a non-uniform flow, presented
by \citet{Maxey&Riley:1983} based on the previous model of \citet{Corrsin&Lumely:1956}.
There are five dominant terms of forces acting on a point-wise particle
in Eq.~\eqref{eq:particle equation of motion}.

\begin{multline}
m_{p}\frac{d\boldsymbol{u}_{p}}{dt}=m_{f}\left(\frac{D\boldsymbol{u}_{f}}{Dt}-\nu\nabla^{2}\boldsymbol{u}_{f}\right)-0.5m_{f}\frac{d}{dt}\boldsymbol{U}_{rel}-6\pi r_{p}\mu\boldsymbol{U}_{rel}-\\
-6\pi r_{p}^{2}\mu\int_{-\infty}^{t}\frac{d\boldsymbol{U}_{rel}/d\tau}{\sqrt{\pi\nu(t-\tau)}}d\tau+(m_{p}-m_{f})\boldsymbol{g}\label{eq:particle equation of motion}
\end{multline}
here $\boldsymbol{u}_{p}$ represents particle velocity, $m_{p}$
is the mass of the particle with radius $r_{p}$ , $\boldsymbol{u}_{f}$
represents the undisturbed fluid velocity at the particle position,
$m_{f}$ the mass of fluid displaced by the particle, $\boldsymbol{U}_{rel}$
represents the relative velocity between the particle and the fluid.
The notation $\frac{d}{dt}=\frac{\partial}{\partial t}+\boldsymbol{u}_{p}\cdot\nabla$
indicates a full (Lagrangian) derivative following the moving particle,
while $\frac{D}{Dt}=\frac{\partial}{\partial t}+\boldsymbol{u}_{f}\cdot\nabla$
represents the material derivative of the moving fluid element. The
terms on the right side of Eq.~\eqref{eq:particle equation of motion}
are the pressure gradient force $\boldsymbol{F}_{p}$, the added-mass
force $\boldsymbol{F}_{a}$, the Stokes drag force $\boldsymbol{F}_{s}$,
the Basset (history or viscous unsteady) force $\boldsymbol{F}_{b}$,
and the buoyancy force $\boldsymbol{F}_{g}$, respectively. This equation
was developed for an isolated, point-like particle in the bulk of
a smooth uniform flow, far from any boundary so that particle-particle
and particle-boundary interactions are excluded. There is also no
lift force term for the point-like particle.

The formulation of \citet{Maxey&Riley:1983} keeps the buoyancy term
but replaces the added mass, Basset, and the Stokes drag terms with
expressions modified by the Laplacian of the local fluid velocity
field, known as ``the Fax\'{e}n corrections'', see for example
\citet{Calzavarini:2009}. Although the equation of motion relies
on the assumption that the particle is much smaller than the smallest
scales of the flow, it was shown that also in cases when the particle
is of the order of the Kolmogorov length scale, the error is not significant
\citep{Bourgoin:2014,Toschi:2009}. \citet{Calzavarini:2012} have
shown that the Fax\'{e}n correction becomes significant for particles
with a diameter of ten times larger than the dissipative length scales
of the flow. In our experiments, the inertial particle size is of
the order of the Kolmogorov length scale. Therefore we use the form
of the terms that does not contain the Fax\'{e}n corrections.

\citet{Sridhar_Katz:1995} reported force measurements on bubbles
in a laminar vortex flow based on the \citet{Maxey&Riley:1983} formulation.
In their study neighboring bubbles were used to estimate the local
fluid velocity. The Basset force and Fax\'{e}n correction were neglected,
the Stokes drag was replaced with a general drag term and the additional
lift force term was introduced. The procedure involved the measurement
of the pressure, inertia and buoyancy forces on each bubble (based
on the interpolation of velocity of the neighboring bubbles) and balancing
their resultant $(\boldsymbol{F}_{T})$ with the lift and drag forces.
Drag and lift forces are therefore the parallel and normal components
with respect to the vector of relative velocity between the particle
and the fluid, $\boldsymbol{U}_{rel}$. Recently \citet{meller:2015}
extended the method of \citet{Sridhar_Katz:1995} for the turbulent
flow case by using two-phase (tracers/particles) experimental data
of \citet{Guala:2008}. In \citet{Sridhar_Katz:1995,meller:2015}
notation, the balance of forces is:

\begin{equation}
\boldsymbol{F}_{T}=\boldsymbol{F}_{D}+\boldsymbol{F}_{L}=-(\boldsymbol{F}_{g}+\boldsymbol{F}_{p}+\boldsymbol{F}_{i}+\boldsymbol{F}_{b})\label{eq:total force}
\end{equation}

where the inertia force is defined for the sake of brevity as follows:

\begin{equation}
\boldsymbol{F}_{i}=m_{p}\frac{d\boldsymbol{u}_{p}}{dt}-\boldsymbol{F}_{a}\label{eq:inertia force1}
\end{equation}
And the drag and the lift are defined as:

\noindent 
\begin{equation}
\boldsymbol{F}_{D}=\frac{\boldsymbol{F}_{T}\cdot\boldsymbol{U}_{rel}}{\left|\boldsymbol{U}_{rel}\right|}\label{eq:drag force-1}
\end{equation}

\begin{equation}
\boldsymbol{F}_{L}=\boldsymbol{F}_{T}-\boldsymbol{F}_{D}\label{eq:lift force-1}
\end{equation}

In the present study we use the method of \citet{Sridhar_Katz:1995,meller:2015},
applied to estimate forces on inertial particles before and during
the lift-off in turbulent flows, based on the particle motion and
the local flow analysis. Since we study the particle lift-off events
of freely moving particles (after their incipient motion), the particle
contact time with the surface is very short, therefore particle-surface
forces, such as adhesion and contact, can be neglected.

The Basset ``history'' force accounts for the viscous unsteady effect
and for the temporal delay in particle response. The delay is due
to the time it takes for the boundary layer on the surface of a particle
to adjust to the varying relative velocity. According to its definition
in Eq.\,\eqref{eq:particle equation of motion}, this term is calculated
as an integral from the the moment of incipient motion to the present
time. Because turbulent flows change rapidly, the earlier moments
contribute a negligible portion and the term can be approximated using
only a few recent time steps. \citet{Mei_Adrian:1992} examined the
inertial effects on the history force and found that for finite particle
Reynolds numbers, the Basset kernel behaves a $t^{-1/2}$ for short
times and decays at a faster rate proportional to $t^{-2}$ at larger
times. More recent studies of \citet{Bombardelli:2008,Lukerchenko:2010,Oliveri:2014}
demonstrated that the Basset force is important for relatively small
particles moving close to solid boundaries, and should be included
in Lagrangian models of bed-load transport if the particle settling
velocity Reynolds number ($Re_{W}=W_{s}d_{p}/\nu$) is below 4000
(where $W_{s}$ is settling velocity, in our case it is $\approx11$
mm/s), and the density ratio $s=\rho_{p}/\rho_{f}$ is of the order
$1-10$. \citet{Ling:2013} have examined the unsteady force terms
and have shown their importance for particles of size comparable with
the Kolmogorov length scale. The authors also studied whether it is
possible to simplify the history integral, and provided a useful criteria
for the choice of a simplified method. Based on the Kolmogorov time
scale estimate (in our case $\tau_{\eta}\approx0.1$ s), and the viscous-unsteady
time scale defined in \citet{Mei_Adrian:1992,Ling:2013} as $\tau_{\mathrm{vu}}=\frac{d_{p}^{2}}{\nu}(\frac{4}{\pi})^{1/3}(\frac{0.75+0.105Re_{p}}{Re_{p}})^{2}$
(in our cases $0.005\div0.07$ seconds) we verify that $\tau_{\mathrm{vu}}<\tau_{\eta}$.
Therefore, we define the Basset viscous unsteady kernel in the following
analysis, in the form of Eq.~\eqref{eq:particle equation of motion},
but integrating for the short interval of $t-0.5\tau_{\eta}$.

\begin{figure}
\noindent \begin{centering}
\includegraphics[width=0.8\textwidth]{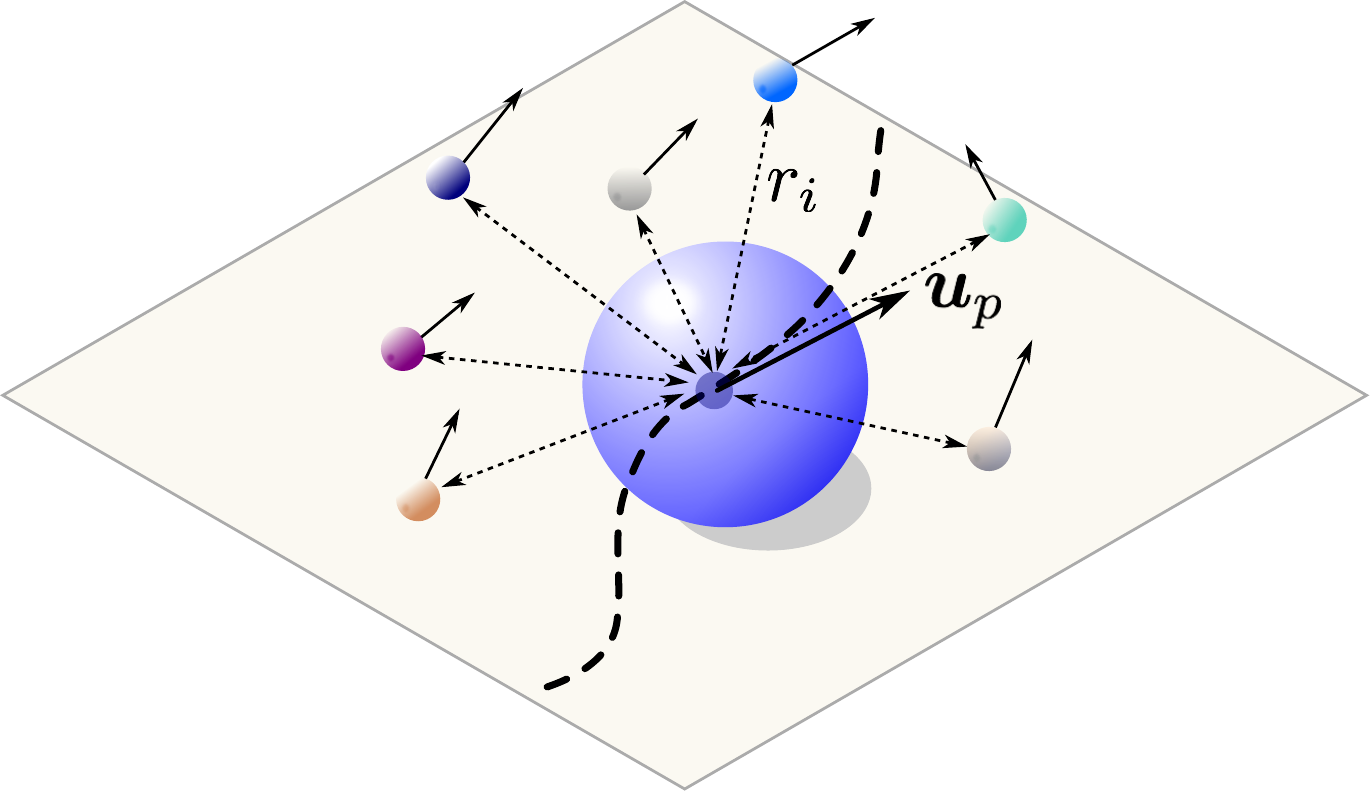} 
\par\end{centering}

\caption{Scheme of definition of the relative velocity $\boldsymbol{U}_{rel}$
between the particle moving at velocity $\boldsymbol{u}_{p}$ and
neighbor tracers, moving at center-to-center distance $\boldsymbol{r}_{i}$.
Dashed line is a schematic trajectory of the particle. \label{fig:Scheme-Urel}}
\end{figure}

The Reynolds number based on the local relative velocity $\boldsymbol{U}_{rel}$
between the particle and local flow ($Re_{r}=\boldsymbol{U}_{rel}d_{p}/\nu$)
at the moment of lift-off was found in the range of $2\div50$. The
schematic definition of the local relative velocity is shown in Fig.\,\eqref{fig:Scheme-Urel}.
A solid particle (large sphere) is moving along the bottom all at
velocity $\boldsymbol{u}_{p}$. Neighbor tracers ($i=1\ldots N$),
move near the solid particle position, distant by distance $r_{i}$.

We implement the same analysis as in \citet{meller:2015} of \emph{inverse
distance weighting} interpolation (sometimes denoted as Shepard interpolation)
see \citet{Shepard:1968}. The interpolated value that represents
the undisturbed velocity at the position of the solid particle is:

\begin{equation}
\boldsymbol{\left\langle u\right\rangle }=\frac{\sum_{i}u_{i}r_{i}^{-p}}{\sum_{i}r_{i}^{-p}}\label{eq:idw}
\end{equation}

where $p$ is the parameter, determined empirically. In this case
the vector of relative velocity is defined through the interpolated
value, and the angular brackets hereinafter define quantities that
are derived similarly:

\begin{equation}
\boldsymbol{U}_{rel}=\left\langle \boldsymbol{u}-\boldsymbol{u}_{p}\right\rangle \label{eq:urel}
\end{equation}

\begin{figure}
\noindent \begin{centering}
\includegraphics[width=0.8\textwidth]{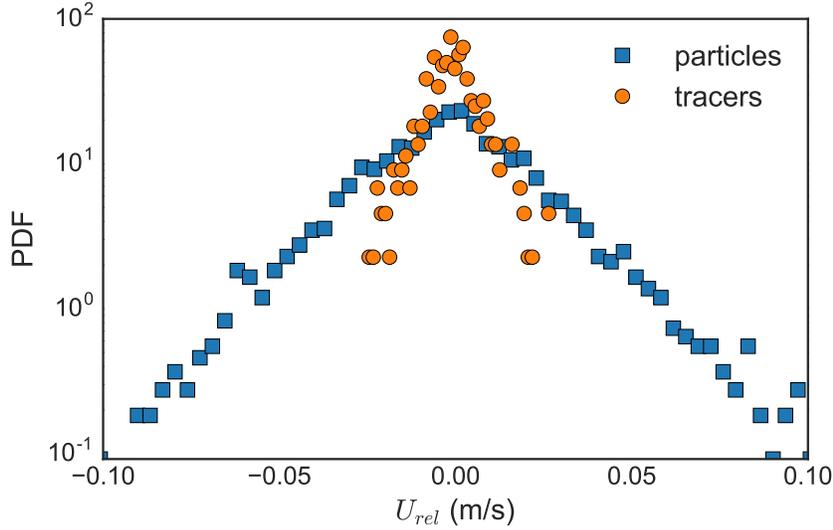} 
\par\end{centering}

\caption{Probability density function of relative velocity components for solid
particles (squares) and flow tracers (circles), estimated through
Eq.\,\eqref{eq:idw}. \label{fig:Comparison-of-relative}}
\end{figure}

Quality of the relative velocity estimate depends on the distribution
of tracers surrounding the particle, namely the number, velocity,
acceleration and distance $r_{i}$. Tracer self-test is a fidelity
measure applicable to our case, similarly to the method used by \citet{Cisse:2013}
to compare the effects on tracers versus inertial particles. In our
case we use the interpolation method at the positions of the tracers
themselves and compare it to the measures at the position of the measured
particles. In the case of tracers \citet{meller:2015} have suggested
to use the statistical self-description index test us a group of tracers
(in all measurements). We repeated this analysis to find the power
exponent $p=3$ to be the optimal choice for the present case. The
probability density function (PDF) of relative velocity $\boldsymbol{U}_{rel}$
for tracers and solid particles is expected to be significantly different.
Such comparison is shown in Fig.~\eqref{fig:Comparison-of-relative}.
The results are similar to those of \citet{Sridhar_Katz:1995,meller:2015}.
It is important to note that the relative velocity magnitude of 1-2
cm/s is one to two orders of magnitude smaller than the flow velocity
in this experiment and points out that our freely moving particles
move along the wall with the velocity only slightly different from
the flow velocity. 

It is important to note that our main goal is to estimate relative
contribution of the force terms before, during after the lift-off
events based on local turbulent flow parameters. Taking into account
the inhomogeneous distribution of flow tracers near the wall, we deal
with an additional trade-off in the selection of the number of nearest
neighbor tracers. We first encountered this problem in our preliminary
study \citet{Traugott:2011} and studied it in more details in \citet{meller:2015}.
Force estimate using Lagrangian data based on Eq.~\eqref{eq:total force}
requires tracers to stay within the particle neighborhood for a certain
time. This is especially important for the Basset force estimate.
Strongly fluctuating number of tracers during the time interval of
particle lift-off event adversely affects the results. We also experience
an optical problem: tracers that are too close to the particle are
being shaded and disappear from tracking database. Following the tracers
self-test and sensitivity analysis of the inverse distance weighting
parameter $p$, we found that the force estimate would not vary significantly
if we include in the interpolation particles that are closer than
$80\eta$ from the center of a solid particle in the horizontal direction
(the diameter of the particles is of the order of the Kolmogorov length
scale, $\eta$), and less than $30\eta$ in the wall-normal direction.
These restrictions act in addition to the inverse distance weighting
method that increases weights of the nearest neighbors and reduces
weights of the tracers beyond 2-3 particle diameters. As an outcome
of these restrictions, at each time step, forces were calculated based
on $20\div100$ tracers. 

We also need to describe in details how the forces are estimated from
the Lagrangian data. Buoyancy force is estimated based on the average
particle radius $r_{p}$ and density $\rho_{p}$ according to Eq.~\eqref{eq:buoyancy force}
and this is the only force component that does not depend on the data:

\begin{equation}
F_{g}=\frac{4}{3}(\rho_{p}-\rho_{l})g\pi r_{p}^{3}\label{eq:buoyancy force}
\end{equation}
Buoyancy acts on the inertial particles downwards (i.e. in the negative
$y$ direction) and on average, taking into account the size distribution
of the solid particles, its magnitude is $5.7\times10^{-8}$ N.

\noindent The pressure and inertia forces were calculated from the
Lagrangian acceleration of the inertial particle and that the local
flow surrounding it, according to Eq.~\eqref{eq:pressure force}
and ~\eqref{eq:inertia force} respectively (see \citet{meller:2015}
for more details):

\begin{equation}
\boldsymbol{F}_{p}=\frac{4}{3}\rho_{l}\pi r_{p}^{3}\left\langle \frac{D\boldsymbol{u}_{f}}{Dt}\right\rangle \label{eq:pressure force}
\end{equation}

\begin{equation}
\boldsymbol{F}_{i}=-\frac{4}{3}\pi r_{p}^{3}\left(\rho_{p}\frac{d\boldsymbol{u}_{p}}{dt}+m_{a}\rho_{l}\left\langle \frac{D\boldsymbol{u}_{p}}{Dt}-\frac{D\boldsymbol{u}_{f}}{Dt}\right\rangle \right)\label{eq:inertia force}
\end{equation}
where $\left\langle \frac{DU_{f}}{Dt}\right\rangle $ is the interpolated
acceleration evaluated at the center of the solid particle according
to Eq.~\eqref{eq:idw}. $\left\langle \frac{Du_{p}}{Dt}-\frac{Du_{f}}{Dt}\right\rangle $
denotes the average difference between particle and tracers accelerations.
The first term in the inertia force is the body force of the particle,
which depends on the mass and acceleration of the particle. The second
term is the apparent mass force. According to literature, for instance
\citet{Van-der-Geld:1991}, the so-called added mass coefficient of
the solid particles is $m_{a}=0.5$. The Basset force was estimated
according to the definition and taking into account the finite time
kernel as shown by \citet{Mei_Adrian:1992}:

\begin{equation}
\boldsymbol{F}_{b}=6\pi r_{p}^{2}\mu\intop_{t-0.5\tau_{\eta}}^{t}\frac{d\boldsymbol{U}_{rel}/d\tau}{\sqrt{\pi\nu(t-\tau)}}\label{eq:basset force}
\end{equation}
which means that the average relative velocity was differentiated
in time to obtain the relative acceleration term $d\boldsymbol{U}_{rel}/d\tau$.
At each time step, the Basset force was calculated using the backward
finite-difference scheme, based on the previous time steps. We have
verified that the Basset force value does not change if we integrate
longer than 1/2 the Kolmogorov time scale.

\noindent The force vector that balances the sum of the buoyancy,
inertia, pressure and Basset force terms is called the total force,
$\boldsymbol{F}_{T}$, according to Eq.\,\eqref{eq:total force}.
Following \citet{Sridhar_Katz:1995,meller:2015}, the total force
vector was decomposed into two orthogonal components to obtain the
magnitude of the drag and lift forces. Drag is the component of total
force parallel to the relative velocity, $\boldsymbol{U}_{rel}$ and
the lift is perpendicular to it, according to Eqs.~\eqref{eq:drag force-1}
and ~\eqref{eq:lift force-1}.

\section{Results and discussion\label{sec:Results-and-discussion}}

\begin{figure}
\begin{centering}
\subfloat[]{\begin{centering}
\includegraphics[width=0.49\textwidth]{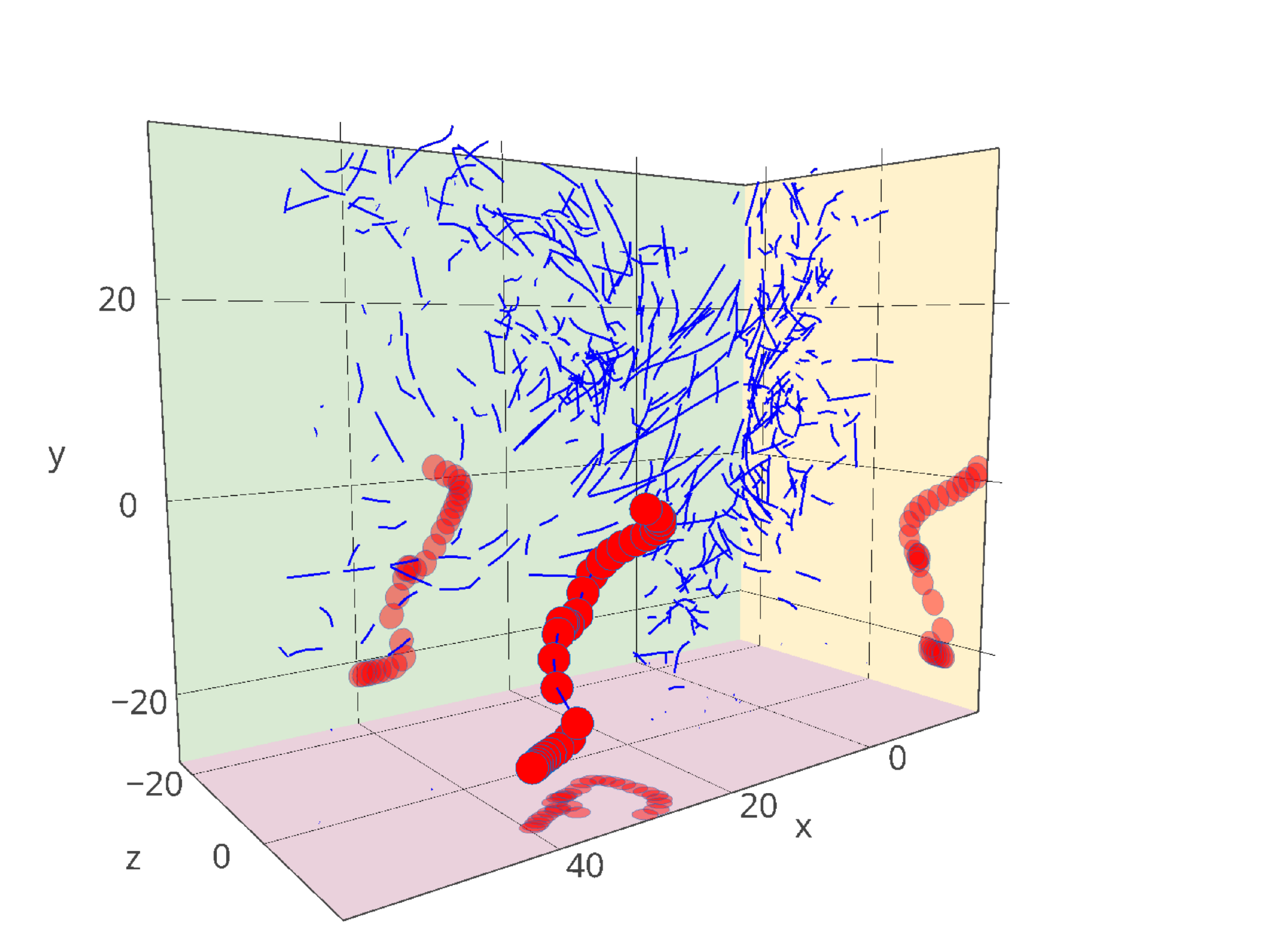} 
\par\end{centering}

}\subfloat[]{\begin{centering}
\includegraphics[width=0.49\textwidth]{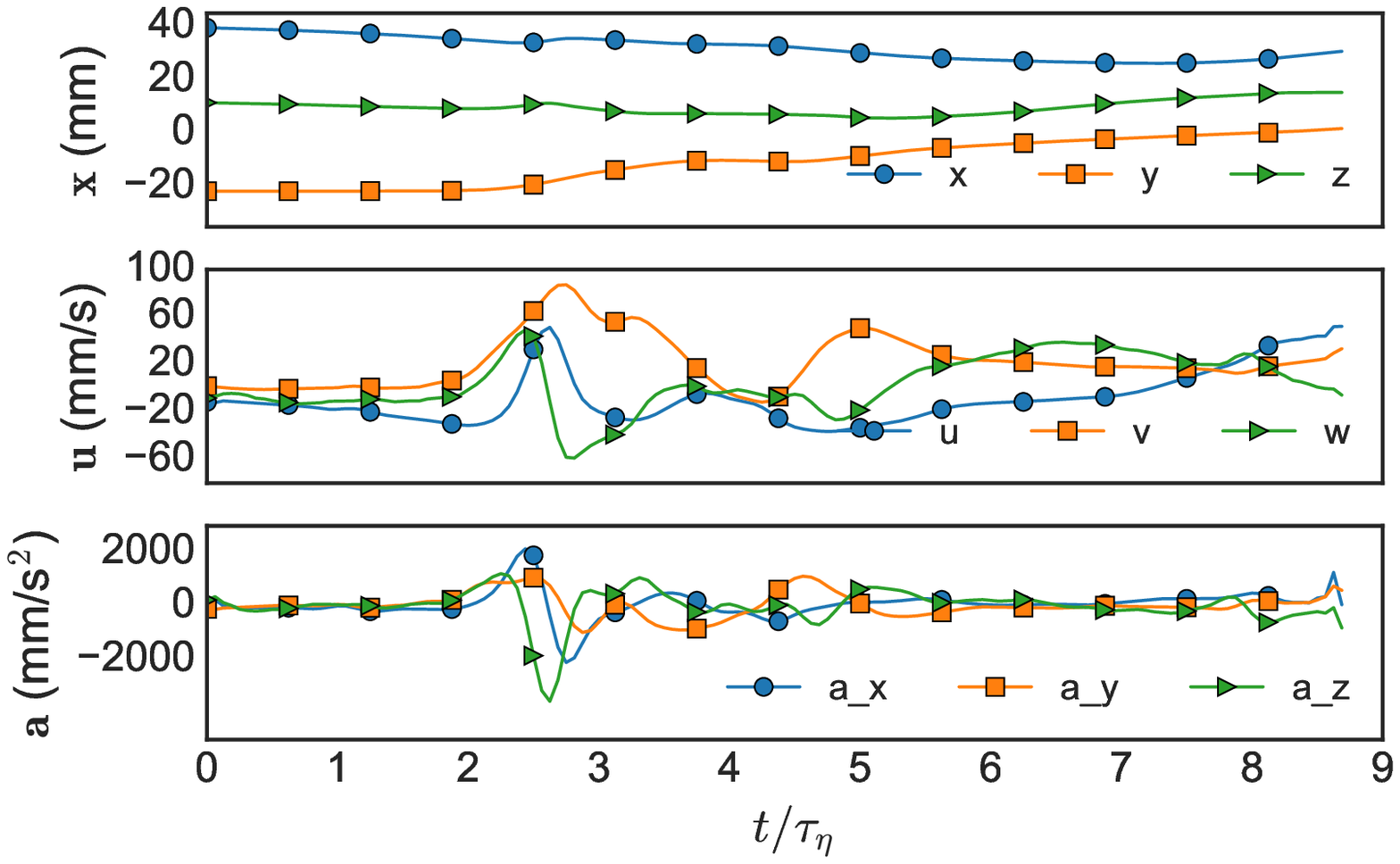} 
\par\end{centering}

}
\par\end{centering}

\caption{(a) Isometric sketch of a Lagrangian trajectory of a lifted particle.
(b) Position, velocity and acceleration components of a particle experiencing
the lift-off event. The frequency in this run was 1.7 Hz. \label{fig: Long trajectories properties Vs time} }
\end{figure}

Particle tracking velocimetry experiments provide the Lagrangian trajectories
defined as locations of particles in time and space, $\boldsymbol{x}_{p}(t)$,
for large particles and $\boldsymbol{x}_{f}(t)$ for flow tracers.
Fig.~\eqref{fig: Long trajectories properties Vs time}(a) presents
a three-dimensional view of an example of Lagrangian trajectory exemplifying
a lift-off event of a large particle as it leaves the wall and moves
upwards, along with numerous trajectories of tracers surrounding the
particle at different time instants. Coordinates are given in millimeters
with respect to the origin predefined by the calibration target, $y$
is the vertical (parallel to gravity) direction. Fig.~\eqref{fig: Long trajectories properties Vs time}(b)
demonstrates a more quantitative analysis of the position, velocity
and acceleration vector components in time of the particle. A combination
of the visual and quantitative information illustrates the complexity
of the problem. At the beginning, for the first two Kolmogorov time
scales, the particle moves along the wall ($y=-21$ mm), as we see
the values change only for $x$ and $z$ components. The particle
is either sliding or rolling on the bottom wall. At approximately
$2\tau_{\eta}$ the lift-off event occurred and the particle resuspended
into the flow. After the lift-off event (after detachment from the
wall) all the velocity and acceleration components change, as expected
for a three-dimensional turbulent flow case. The plot also underlines
the intrinsic difficulty to define the precise time instant of the
lift-off event as the vertical position, velocity and acceleration
components do not change instantly. The definition of the detachment
can be based on either position, velocity or acceleration, as we discuss
in the following.

\subsection{Definition of the lift-off event\label{sub:Defining-the-lift-off}}

We distinguish the moment of lift-off from the typically defined incipient
motion, which is the beginning of particle movement on the bottom
surface from a fixed position. Although we could track particles that
at some moment of time have zero velocity, this experiment is difficult
to perform on freely moving particles. We focus on the lift-off of
the freely moving particles but we also include the events of the
particles that are lifted with a negligible horizontal velocity. The
moment of lift-off and the influence of the local flow characteristics
and fluid-particle forces on the event and particle motion is our
major focus.

\begin{figure}
\begin{centering}
\includegraphics[width=0.8\textwidth]{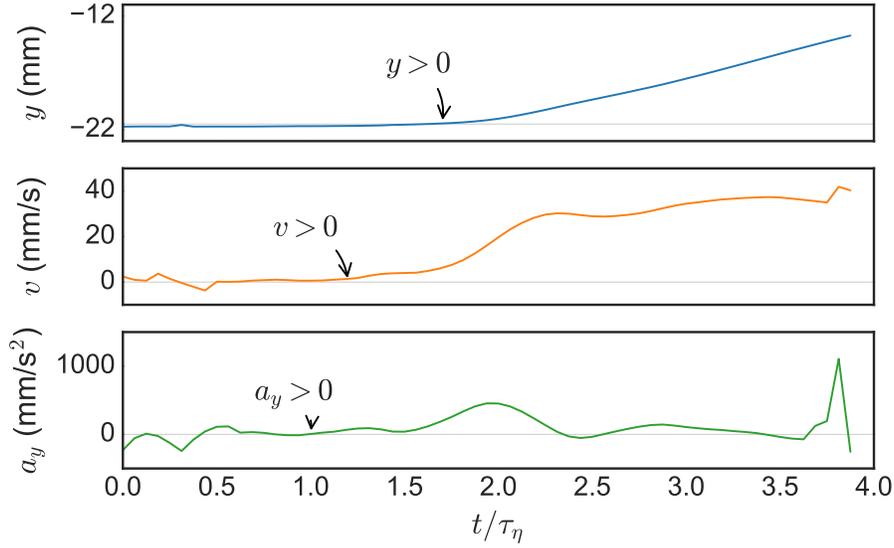} 
\par\end{centering}

\caption{Vertical components of position, velocity and acceleration of a resuspended
particle during the lift-off event, extracted from the run at 1.5
Hz.\label{fig: type2_event_definition_inquiry_22}}
\end{figure}

A standard lift-off event is defined as the detachment of a particle
from the wall, i.e. breaking the contact with the surface. Zooming
into the data to visualize the event, as shown in Fig.~\eqref{fig: type2_event_definition_inquiry_22},
we identify the predictable scenario: first, at one Kolmogorov time
scale from the arbitrary time instant defined as $t_{0}$ the acceleration
component $a_{y}$ changes to positive and its value increases monotonically;
after additional $1/5\tau_{\eta}$ (about $20$ milliseconds), the
vertical velocity value changes (other velocity components change
continuously as shown in Fig.~\eqref{fig: Long trajectories properties Vs time})
and only after additional $1/3\tau_{\eta}$ (30 msec) the $y$ coordinate
changes and the particle detaches from the wall. It means that one
could define the moment of lift-off by either the first moment of
the positive acceleration, positive velocity or the last moment of
$y=y_{wall}$. In addition, one has to take into account the experimental
uncertainty, thus the definition has to be dependent on some time
or magnitude threshold.

Although the time interval of the lift-off event is shorter than one
Kolmogorov time scale, we emphasize that the definition of the moment
of the event based on one of the properties might affect the analysis
and the results. The time point for lift-off in this study is defined
in this study as the first time step in which the position of the
particle in $y_{p}>y_{wall}+0.1$ mm, disregarding changes in position
smaller than the 1/5 of the particle diameter (small jumps during
the horizontal motion along the wall). In order for the event to be
considered as a lift-off event, the particle must remain in suspension
for at least 1/2 Kolmogorov time scale after the detected moment.
If a particle returned to the bottom and remained on the wall for
at least another 1/2 Kolmogorov time scale, a separate lift-off event
of the same particle could be identified. As we explained in the Section
\eqref{sec:Materials-and-methods}, out of many resuspension events,
only those that were within the field of view of the 3D-PTV system,
successfully tracked during the full length of the event, got lifted
and not land immediately afterward and have sufficient tracer density
surrounding the lift-off event, were considered for the following
analysis.

\subsection{Forces on particles before and during lift-off events \label{sub:Forces-on-particles}}

\begin{figure}
\begin{centering}
\subfloat[]{\includegraphics[width=0.49\linewidth]{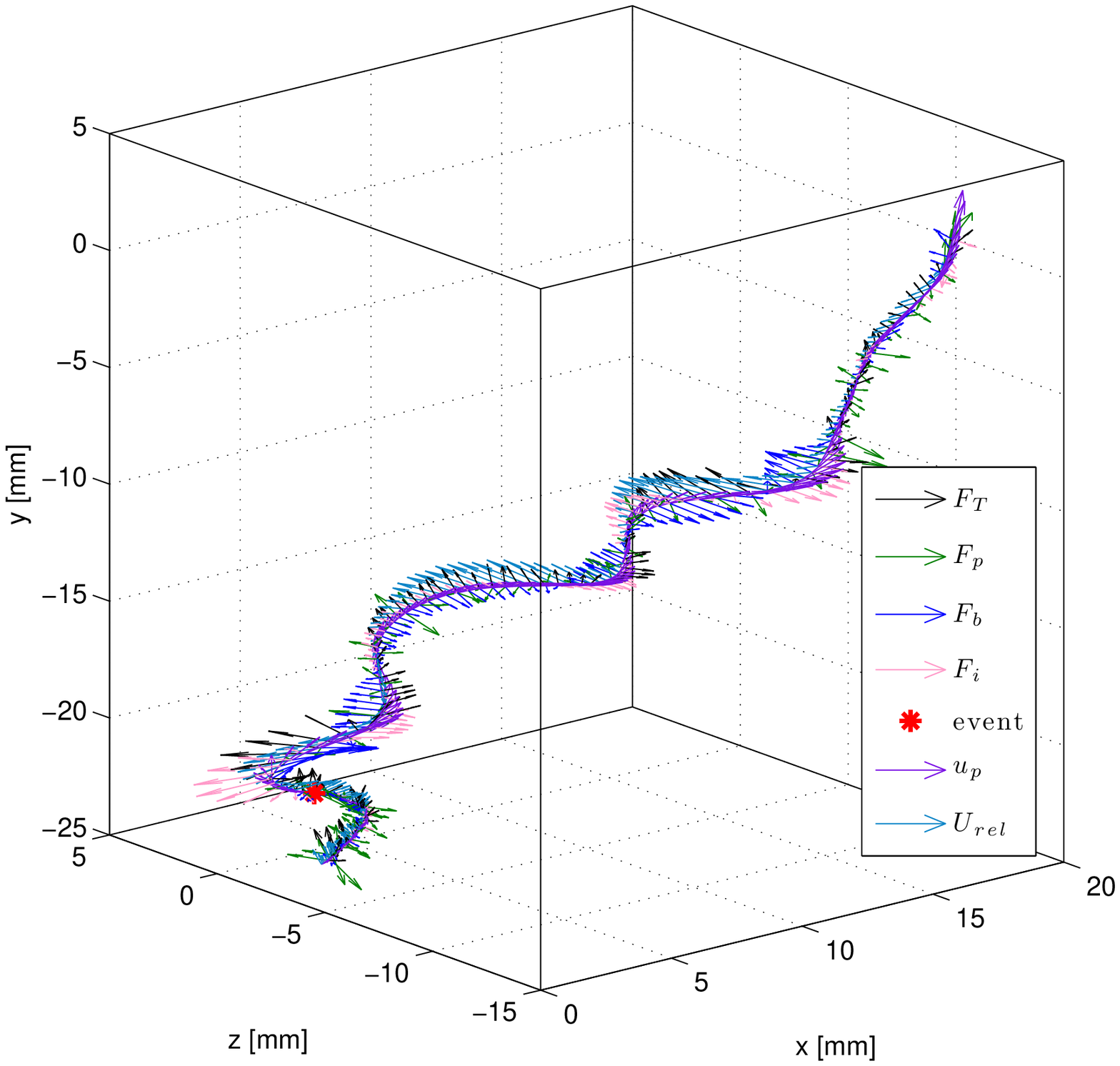}

} \subfloat[]{\includegraphics[width=0.49\linewidth]{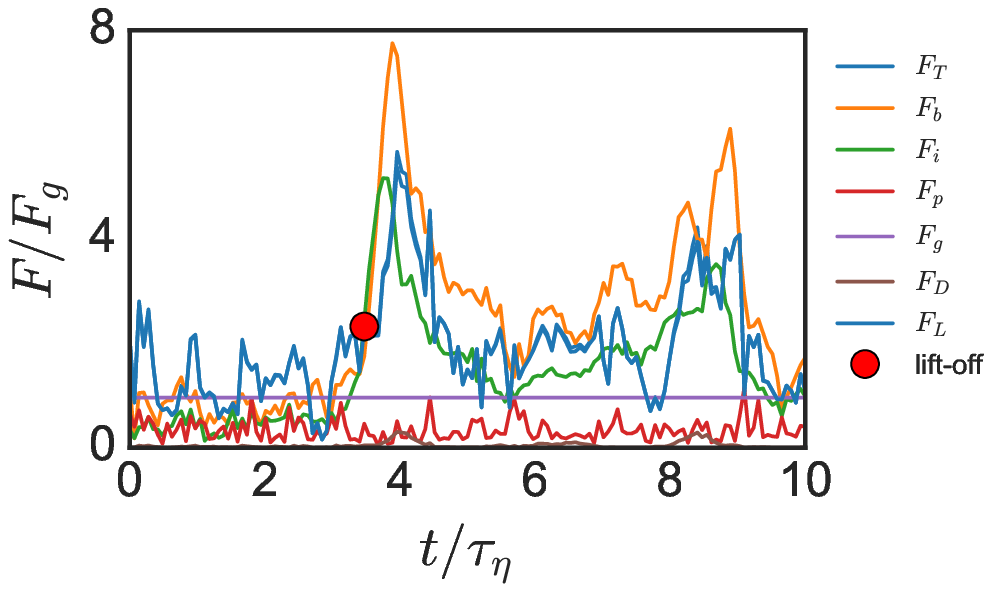}

}
\par\end{centering}

\caption{\label{fig:trajectory of a re-suspended particle with forces}(a)
presents an isometric sketch of a Lagrangian trajectory of a particle
demonstrating the magnitude and direction of velocities and forces
acting on the particle at each time step, extracted from the run at
frequency of $1.7\ \mathrm{Hz}$. (b) Magnitude of the forces acting
on the particle in time, moment of lift-off is marked as the red point. }
\end{figure}

Forces acting on the particles during lift-off events were estimated
according to the methods described in Section~\eqref{sub:Forces-on-particles}.
Fig.~\eqref{fig:trajectory of a re-suspended particle with forces}(a)
exemplifies a Lagrangian trajectory of a lifted-off particle, demonstrating
the relative magnitude and direction of the forces acting on the particle,
particle velocity $(\boldsymbol{u}_{p}$) and the relative velocity
$(\boldsymbol{U}_{rel}$) between the particle and local fluid at
each time step: before, during and after lift-off. Fig.~\eqref{fig:trajectory of a re-suspended particle with forces}(b)
demonstrates the magnitudes of the forces acting on the particle in
time normalized by the magnitude of buoyancy force. The moment of
lift-off is marked by the red point. We observe that the force terms,
excluding drag, are of similar magnitude before the lift-off event.
The drag force is much smaller as compared to other terms. The magnitude
of lift, inertia and Basset forces increase about 4-10 time steps
(1/4-2/3$\tau_{\eta}$) right before the lift-off and remain high
until 10 time steps (2/3 $\tau_{\eta}$) after the lift-off event.
As the particles are fully entrained, they move almost like the tracer
particles in a strong turbulent flow and the relative velocity sharply
decreases, as was estimated by \citet{meller:2015}. Here we focus
only on the relative magnitude of the different terms of forces before
and during the lift-off with an emphasis on the magnitude of Basset
force. The magnitude of Basset force is approximately one-half of
lift force value before the lift-off event, and shows the largest
values at and after the lift-off. The magnitude of the drag force
is the least important in this type of resuspension events, approximately
two orders of magnitude lower than the lift force. This finding is
in apparent contradiction with the large body of literature that emphasizes
the drag force over lift force. It is explained by the fact that the
particle is already moving (either sliding or rolling) at the velocity,
which is close to that of the flow velocity in the near wall region.
The lift force, as it is understood here, is caused by the local shear.
The Basset unsteady-viscous force seems to be also non-negligible
and contributes to the lift-off events. We also observe that $F_{T}$
is almost completely perpendicular to the relative velocity vector,
such that the magnitudes of $F_{T}$ and $F_{L}$ are overlapping.
Due to relatively large experimental uncertainty (as estimated in
the Appendix according to standard error propagation methods) we are
not able to claim the precise values of the forces acting on the particles.
Nevertheless, the estimate of errors enforce that we can safely compare
the magnitudes of the force components. The largest uncertainty is
for the inertia force ($\varepsilon_{F_{i}}=1.2\cdot10^{-7}\ \mathrm{N}$
) as its value is determined by subtracting acceleration vectors.
Consequently, the accuracy of drag and lift values is also limited,
but it is sufficient to emphasize that the lift is at least an order
of magnitude higher than the drag force.

\begin{figure}
\includegraphics[width=1\linewidth]{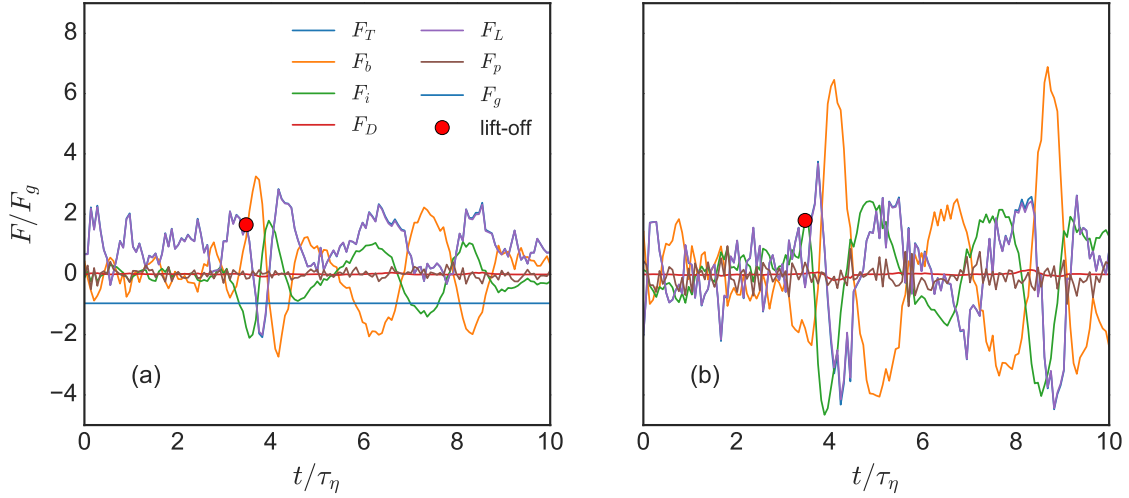}

\caption{\label{fig:re-suspended particle with forces_vertical_horizontal}
Time history of forces acting on the particle, obtained for the grid
frequency of $1.7\ \mathrm{Hz}$, and the moment of lift-off is marked
by the red point. (a) vertical components (b) horizontal components.}
\end{figure}

One attempt to reduce the experimental uncertainty, yet preserve the
insight from the measured data, is to decompose the forces differently.
Instead of using a projection on the vector of relative velocity that
changes abruptly in the turbulent flow near the wall, we project them
on the vertical (gravity) and horizontal planes. We normalize the
force components using the buoyancy force magnitude and show the horizontal
and vertical components in in Fig.\ \eqref{fig:re-suspended particle with forces_vertical_horizontal}.
The buoyancy force is negative in $y$ (vertical) direction and appears
as a straight line in the normalized plot in Fig.\ \eqref{fig:re-suspended particle with forces_vertical_horizontal}
(a). In the horizontal direction, magnitude of all forces (except
for the drag) before lift-off are similar, while in the vertical direction,
magnitude of Basset and lift before lift-off are higher than pressure
and inertia. The event is marked by the point in which the sum of
the lift force and the Basset force is high. 

\begin{figure}
\includegraphics[width=1\linewidth]{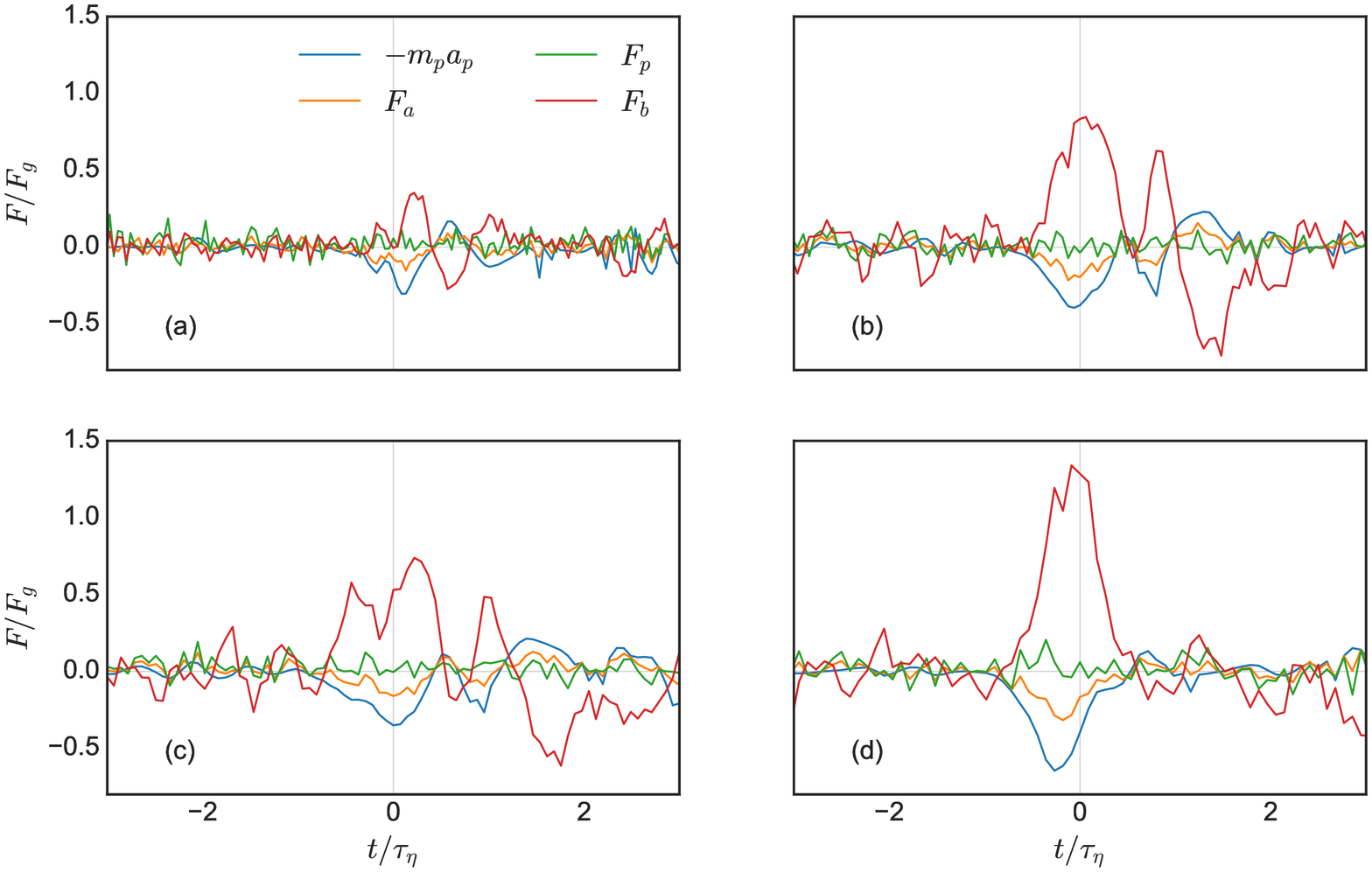}

\caption{\label{fig: vertical pressure first and second inertia}Vertical components
of the first ($-m_{p}\boldsymbol{a}_{p}$) and second term (apparent
mass force, $\boldsymbol{F}_{a}$) of the inertia force, pressure
$\boldsymbol{F}_{p}$ and Basset force, $\boldsymbol{F}_{b}$ . Curves
represent the ensemble averaged quantities, averaged over all trajectories
at the same frequency. Panels (a-d) correspond to the frequencies
1.5 - 2.1 Hz, respectively. }
\end{figure}

The inertia force equation includes two terms: the first is the body
force of the particle ($-m_{p}a_{p})$, and the second is the apparent
mass force. Fig.~\eqref{fig: vertical pressure first and second inertia}
presents the vertical components of body, apparent mass, pressure
and Basset forces, for two different frequencies of the oscillating
grid (a,b). Every line represents the ensemble averaged magnitude
of the force of all the particles suspended and successfully tracked
at the same frequency within the time interval of the experiment (20
and 8 particles for frequencies of $1.7$ and $1.9\ \mathrm{Hz}$
respectively). Eight time steps (1/2 $\tau_{\eta}$) before lift off,
the vertical component of body force ($-m_{p}a_{p}$) decreases below
zero, indicating a consistent increase of particle vertical acceleration
values. This is also the time instant at which the vertical component
of the body force and vertical components of apparent mass and pressure
forces are separated from one another. The difference between them
increases until the lift-off event. Later the vertical acceleration
decreases and the difference between the forces also decrease. Since
the pressure force depends only on fluid acceleration and the apparent
mass force depends both on the fluid and the particle response, we
interpret the situation as follows: during the lift-off event the
particle moves faster than the surrounding flow in the vertical direction,
moving away from the surface. As the particle moves into a faster
turbulent flow in the bulk, the velocity difference between particle
and flow decreases. We also note that about $0.5\div0.7\,\tau_{\eta}$
before the lift-off, the value of the vertical Basset force component
increases significantly and remains high after the lift-off event.
The same trend was observed for all particles and all frequencies,
shown in Fig.\,\eqref{fig: vertical pressure first and second inertia}(a-d).

We carefully examined the force traces of the particles that were
lifted-off at different frequencies, different initial positions and
initial velocities. We were looking for a consistent change in time
that can reveal the necessary and sufficient local flow conditions
that cause a lift-off event. The changes in frequencies have changed
the local flow Reynolds number, and definitely changed the probability
of the lift-off events. However, the local flow conditions that are
responsible for the lift-off event are presumably unrelated to the
global flow changes. The lift-off events we observed in our experiment
are seemingly random events without any cycling or specific time scale
associated with it. The only consistent and repetitive observation
is of increasing Basset force (a sort of accumulative increase of
relative velocity) right before the lift-off events. This observation
is consistent and appears in the ensemble averaged manner for all
frequencies, as shown in Fig.~\eqref{fig: vertical pressure first and second inertia}.
It could imply, that the observed peaks of the Basset force are the
possible triggers of the lift-off event.

\section{Summary and conclusions\label{sec:Summary-and-conclusions}}

In the present study we obtained experimentally the fluid-related
forces acting on inertial particles that are free to move on a smooth
surface. We focused on the lift-off events of suspending particles
into a turbulent flow without significant mean shear, unlike the wind
tunnel and open water channel flow experiments. We have measured the
particle trajectories along with the local flow conditions, before,
during and shortly after the lift off events. In order to simplify
the task of tracking fast moving particles, we worked in an oscillating
grid chamber, typically used for resuspension studies. This allows
to repeatedly measure the lift-off events of the same particles that
are suspended and deposited, providing a large data set from which
the events can be selected during the post-processing. Unique, direct
simultaneous measurements of local flow conditions and 3D Lagrangian
trajectories of the inertial particles, before and during lift-off,
were performed using a 3D-PTV technique for several frequencies of
the grid. The method of \citet{Sridhar_Katz:1995} that was used to
measure forces on bubbles in a laminar vortex was recently extended
for the turbulent flow conditions by \citet{meller:2015} and used
to compare the resuspension from smooth and rough surfaces by \citet{Shnapp:2015}.
Utilizing this method, the pressure, inertia and the Basset history
forces were calculated from the Lagrangian acceleration of the inertial
particle and the surrounding flow tracers.

We can summarize the main findings as follows. The magnitudes of all
forces, excluding the drag force (which is much weaker than other
terms) immediately before the lift-off event are similar. The values
are close to the magnitude of the buoyancy force which we know from
the particle size and density. The Basset history force (understood
as a viscous unsteady force) is found to be non-negligible. Its magnitude
is approximately one-half of that of the lift force before the lift-off
event, and highest of all forces immediately after the lift-off event.
We can conclude that this term, often neglected in numerical modeling
and analysis of experimental results, has to be included in the description
of the problem of resuspension of freely moving particles, especially
in the liquid-solid case. Our measurements are in very good agreement
with recent investigations that took a numerical approach \citep{Bombardelli:2008,Lukerchenko:2010,Oliveri:2014}
as they fall into the same range of parameters, i.e. $Re_{w}=W_{s}d/\nu<4000$
and $s=\rho_{p}/\rho_{f}\approx1$.

Clearly these findings do not necessarily extrapolate to the cases
of particles much heavier than the fluid. In order to understand the
contribution of the Basset force to the lift-off (and possibly to
the incipient motion in general), additional studies are required;
for instance, different particle to fluid densities ratios, $s\gg1$,
such as gas-particle flows or addition of the mean streamwise velocity
and the effect of boundary layers.

Furthermore, the distributions of forces acting on inertial particles
in this type of a turbulent flow indicate that the magnitude of lift
is dominant in this case. In addition, as the magnitude of the total
force measured from particle acceleration $F_{T}$ and lift forces
are very similar, the lift force can be estimated indirectly from
much simpler experiments that do not involve two-phase tracking.

While searching for the necessary and sufficient flow conditions that
precede or define the lift-off event, we have found a peculiar persistent
increase of Basset force before the lift-off. This fact points to
a possible trigger required for lift-off event to occur. Our results,
although performed in different settings, could be in some sense related
to the recent findings of \citet{celik:2010}, among others. The authors
suggested that the time interval of the force acting on the particle
is more important that its value. It is possible that our important
finding of the Basset force relates to the same viscous unsteady effects
that are all related to the establishment of the boundary layer and
the wake of the particle. Other studies have already mentioned the
importance of the pressure gradient in mobilizing particles, for instance
of \citet{Schmeeckle:2007,Zanke:2003}. We arrived to a similar conclusion
observing the increase of the Basset force.

It is important to remember that we do not resolve the flow close
enough to the particle in order to claim that we can accurately estimate
the magnitude of the force. The experimental trade-off could lead
to overestimated or underestimated values. The important part, nevertheless,
is the comparative study of the force terms that are derived with
a similar experimental uncertainty. In order to get a better view
of the proposed mechanisms, a much higher resolution of the instantaneous
flow fields and additional pressure gradient measurements below and
above the test particles, before and during the moment of lift-off,
are required. Such an experiment, however, seems not to be feasible
in the visible future as it will require a super-resolution 3D-PTV,
along with the pressure-sensitive painting or miniature pressure catheters
embedded in the flow at the place and time of particle resuspension.
It is possible that the accurate numerical simulations that resolve
both the large and the small scales and track the solid particles
in the flow, could provide a supportive evidence. We also hope that
the obtained results can be helpful for the better predictions in
the fields of sediment transport, pneumatic conveying, fluidized bed
and bio-reactor design.

\section*{Acknowledgments}

The authors are thankful to Mark Baevsky for the technical assistance
with the experiments.

 \bibliographystyle{plainnat}
\bibliography{bibliography}

\pagebreak{}

\appendix
\section{Particle Tracking Velocimetry and uncertainty analysis}
\subsection{Post-processing of 3D-PTV}

The main tool in our work is the three-dimensional particle velocimetry
(3D-PTV), explained in details in a collection of works, for instance
\citet{Papantoniou:1993a,Papantoniou:1993b,Dracos:1996}. The work
of \citet{Luthi:2005} also provides a careful error analysis of velocity
and velocity derivatives data. For the sake of clarity we briefly
review here the main steps and in the following the errors and uncertainty
analysis. It is noteworthy that we use the implementation of the spatial-temporal
algorithm that increases the tracking efficiency and allows to work
with particle seeding densities high enough to measure velocity derivatives.

3D-PTV comprises of two major steps (after the calibration part that
is the main source of the bias error, described below): \emph{i})
determination of particle positions in space, and \emph{ii}) tracking
(linking) of tracers (or particles) in consecutive images. The spatial-temporal
algorithm exploits the redundant information in image (pixels) and
space (mm) of several snapshots, reducing the error of particle positions
in time, $x(t)$. Tracking step involves the a) search in the radius
adapted using the maximum velocities, b) limited Lagrangian acceleration,
c) heuristic choice of the trajectory after 4 consecutive time steps
with the total minimal Lagrangian acceleration. The raw data in a
form of linked positions of tracers/particles are processed using
a low-pass, moving cubic polynomial filter \citet{Luthi:2005}, providing
their filtered positions, velocities and accelerations. The filter
is defined as

\begin{equation}
\widehat{x}(t)=c_{0}+c_{1}t+c_{2}t^{2}+c_{3}t^{3}
\end{equation}
The expression is fitted to 21 points in time, from $t-10\Delta t$
to $t+10\Delta t$, for each component ($x,y,z)$ at each time step,
$t$ . The filtered velocities and accelerations are defined from
the polynomial coefficients:

\begin{eqnarray*}
\widehat{u}(t) & = & c_{1}+2c_{2}t+3c_{3}t^{2}\\
\widehat{a}(t) & = & 2c_{2}+6c_{3}t
\end{eqnarray*}

Using filtered velocities, $\widehat{u}(x,t)$, spatial and temporal
derivatives could be interpolated for every particle trajectory point.
If the seeding density is sufficient, the linear approximation is

\begin{equation}
\widehat{u}(x_{0})=c_{0}+c_{1}x+c_{2}y+c_{3}z
\end{equation}

where

\[
c_{1}=\frac{\partial u}{\partial x},\quad c_{2}=\frac{\partial u}{\partial y},\quad c_{3}=\frac{\partial u}{\partial z}
\]
In theory, four neighbors are sufficient to solve the equation and
estimate the four coefficients, $c_{i}$. In reality, the errors in
filtered velocities lead to a large error propagation. Therefore,
we use over-determined linear system with information from $n>4$
points, that is solved using the least-square method:

\begin{equation}
\boldsymbol{A}c=u,\quad c=\left(\boldsymbol{A}^{T}u\right)\left(\boldsymbol{A}^{T}\boldsymbol{A}\right)^{-1}
\end{equation}

where $\boldsymbol{A}$ is the matrix of positions. Similarly the
temporal derivative is obtained, through the ansatz:

\begin{equation}
\widehat{u}(x_{0})=c_{0}+c_{1}x_{1}+c_{2}x_{2}+c_{3}x_{3}+c_{4}t
\end{equation}

In this case, the time span for the time derivative is 4 time steps,
$4\Delta t$ and the estimate is the least-squares central difference
gradient method with a residual error of the order of $\mathcal{O}(\Delta t)^{2}$.
We suggest the reader to refer to \citet{Luthi:2005} the full description
of the method and careful error analysis of a general 3D-PTV experiment
and data processing that leads to the high order estimates. The tests
and checks standard and embedded in our post-processing software \citet{openptv}
verify that the local analysis fulfills the relative divergence $\delta\leq0.1$,
defined as

\begin{equation}
\delta=\frac{\left|\partial u/\partial x+\partial v/\partial y+\partial w/\partial z\right|}{\left|\partial u/\partial x\right|+\left|\partial u/\partial y\right|+\left|\partial u/\partial z\right|}
\end{equation}

The errors of Lagrangian accelerations, $a=Du/Dt$ , are defined directly
from the measurements along particle trajectories (see below the quantitative
analysis). The components of accelerations, namely local acceleration
$a_{l}=\partial u/\partial t$ and convective acceleration, $a_{c}=u_{j}\partial u_{i}/\partial x_{j}$
are estimated using the locally interpolated velocities and velocity
derivatives, as described above. It is important to note that both
errors are significantly larger than the one of the Lagrangian, full
material derivative, yet the values of the local and convective acceleration
in turbulence are also order of magnitude larger than the magnitude
of Lagrangian acceleration. Nevertheless, \citet{Luthi:2005} have
compared the theoretical analysis with the empirical values, to show
that the conservative estimate of the relative error is $\epsilon_{a_{c}}\approx20\%$.

\subsection{Uncertainty analysis of the experiment}

Uncertainty analysis starts from a review of the error sources in
the measurements.

The independent variable that determines the 3D-PTV accuracy is the
three-dimensional (3D) particle coordinate in the laboratory frame
of reference, $x,y,z$. Every particle has its own identity and its
trajectory is defined by its position at each time step. Without going
into details of the 3D-PTV method, there are two sources of error:
a) random, precision-type errors due to the image processing algorithm
that detects the center of a particle in multiple camera views, and
b) systematic error due to 3D calibration of the imaging system.

The standard image processing routines provide precision of $0.05\ \mathrm{pixel}$
for the particle centroid. However for realistic images of imperfectly
spherical particles and illumination non-uniformity we estimate the
positioning error of $0.2\ \mathrm{pixel}$ \citet{Dracos:1996}.

Calibration errors arise from several sources: residual lens aberrations,
imprecision in calibration target production and incorrect positioning
of the calibration target. Stationary calibration process uses a high-precision
machined calibration target with numerous target points outnumbering
the number of equations of the photogrammetric model. The software
iteratively solves the error minimization problem of the over-determined
system and estimates the camera interior and exterior parameters.
During the calibration process, the differences between the known
points on the calibration target and the computed points are estimated.
In the presented experiment, the errors in the calibration were estimated
as $\varepsilon_{x}=\varepsilon_{y}=\pm10\ \mathrm{\mu m}$ in the
direction parallel to the imaging plane and $\varepsilon_{z}=\pm30\ \mathrm{\mu m}$
in the direction parallel to the imaging axis (the magnification ratio
in this experiment was $16\ \mathrm{pixel/mm}$). In order to improve
the calibration error, especially in the out-of-plane direction, we
have performed two additional steps: \emph{i}) configuration of the
cameras from both sides of the tank improves the error due to the
intersection of imaging axis of the cameras from two sides (positive
and negative $z$ values; \emph{ii}) additional, dynamic calibration
of the system using the dumbbell (wand) method \citet{openptv}\textbf{.
}We estimate the overall positioning error to be of the order of $\epsilon_{x}=\pm20\mu\mathrm{m}$.

Particle velocity is derived along the particle trajectory using finite
difference of the particle positions in time, for instance one component
is $u=\Delta x/\Delta t$. Particle tracking is performed with the
high frame rate sampling (about 5-10 time steps per Kolmogorov time
scale). The high time sampling allows to filter the positioning random
errors (noise) using the low-pass filter \citet{Luthi:2005}. Thus
velocity and acceleration signals are derived as first and second
derivatives of the filtered trajectories. Considering that the velocity
and acceleration of the particles changes at time steps equivalent
to Kolmogorov time scale of the measurements ($\tau_{\eta}=0.07-0.1\ \mathrm{s}$),
the time interval for filtered velocity and acceleration error calculations
can be estimated as $\triangle t=\frac{\tau_{\eta}}{5}=\frac{0.07\ \mathrm{s}}{5}=0.014\ \mathrm{s}$.
The time delay is controlled by a synchronization unit with very high
precision, therefore has negligible effect on the overall accuracy.
Therefore velocity and acceleration errors were estimated as:

\begin{equation}
\varepsilon_{u}=\sqrt{2\left(\frac{\varepsilon_{x}}{\triangle t}\right)^{2}}\approx2\ \mathrm{mm/s}
\end{equation}

\begin{equation}
\varepsilon_{a}=\sqrt{2\left(\frac{\varepsilon_{u}}{\triangle t}\right){}^{2}}\approx200\ \mathrm{mm/s^{2}}
\end{equation}

Using the uncertainty propagation from the vector components to the
vector magnitude, we obtain the following error estimates for the
flow tracers (denoted with a subscript $_{f}$ ) and particles ($_{p}$),
velocity and acceleration.

\begin{table}
\begin{tabular}{|c|c|c|c|}
\hline 
$\left|u_{f}\right|$  & $\varepsilon_{\left|u_{f}\right|}$  & $\left|u_{p}\right|$  & $\varepsilon_{\left|u_{p}\right|}$\tabularnewline
\hline 
\hline 
0.11  & 0.002 & 0.1  & 0.002\tabularnewline
\hline 
\end{tabular}%
\begin{tabular}{|c|c|c|c|}
\hline 
$\left|a_{f}\right|$  & $\varepsilon_{\left|a_{f}\right|}$  & $\left|a_{p}\right|$  & $\varepsilon_{\left|a_{p}\right|}$\tabularnewline
\hline 
\hline 
7.0  & 0.2 & 2.1  & 0.2\tabularnewline
\hline 
\end{tabular}

\noindent \centering{}\caption{Estimated flow (tracers) and particles velocity and acceleration errors,
(m/s, m/s$^{2}$).\label{tab:estimated particle and tracers velocity error}}
\end{table}

\subsection*{A.3.1 Force estimates uncertainty analysis}

The buoyancy force depends only on the measurements of the particle
density, $\rho_{p}$, fluid density $\rho_{f}$ and the particle radius
$r_{p}$,

\begin{equation}
F_{g}=\frac{4}{3}(\rho_{p}-\rho_{f})g\pi r_{p}^{3}\label{eq:buoyancy force app}
\end{equation}
The error depends on $\rho_{p}$ and $r_{p}$ and can be evaluate
as follow:

\begin{equation}
\varepsilon_{F_{g}}=\sqrt{\left(\frac{\partial F_{g}}{\partial\rho_{p}}\varepsilon_{\rho_{p}}\right)^{2}+\left(\frac{\partial F_{g}}{\partial r_{p}}\varepsilon_{r_{p}}\right)^{2}} 
\end{equation}
Estimated buoyancy force amplitude is $5.7 \times 10^{-8}$ N, the error value is $=2.6 \times 10^{-9}$ N, therefore the relative error is 4.5\%.

\bigskip{}

The pressure force is estimated using the fluid material derivative
along particle trajectory

\begin{equation}
F_{p}=\frac{4}{3}\rho_{f}\pi r_{p}^{3}\left\langle \frac{Du_f}{Dt}\right\rangle \label{eq:pressure force app}
\end{equation}
where $\left\langle \frac{Du_f}{Dt}\right\rangle $ is the volume
average of the Lagrangian acceleration of neighbor tracers. The region
that defines the neighborhood is not well defined. It is a compromise
between the sufficiently high number of tracers used for statistics
and the sufficiently small radius of averaging. In any case the concept
of coarse-grained fluid flow information is used, following \citet{Luthi2007},
among others. In this approach the coarse-grained quantities are defined
as volume integrals:

\begin{equation}
\langle u\rangle=\frac{1}{\forall}\int_{\forall}u(x+x')d^{3}x'
\end{equation}
\citet{Luthi2007} have shown the convergence of the statistics of
the coarse-grained quantities that depends mainly on the seeding density
of the tracers. The minimum number of particles was found to be 12 $\div$ 15
in order to achieve a convergent statistics. The test is the value
of the coarse-grained (or volume averaged) quantity of interest at
monotonically increasing volume of averaging volume, $\forall$. In
our experiment the size of the volume corresponds to a sphere of $25\ \mathrm{mm}$.
The error in the mean Lagrangian acceleration of tracers was estimated
by the standard deviation of tracers acceleration and the number of
tracers. The number of tracers was chosen as the lowest number of
tracers observed at a specific time step in order to obtain the highest
possible error; $\varepsilon_{\langle a_{f}\rangle}=200\ \mathrm{\mathrm{mm/}s^{2}}$,
$\langle a_{f}\rangle=7000\ \mathrm{m\mathrm{m/s^{2}}}$. Therefore,
the estimated error of pressure force is:

\begin{equation}
\varepsilon_{F_{p}}=\sqrt{\left(\frac{\partial F_{p}}{\partial r_{p}}\varepsilon_{r_{p}}\right)^{2}+\left(\frac{\partial F_{p}}{\partial\overline{a_{f}}} \varepsilon_{\left\langle a_{f}\right\rangle }\right)^{2}}
\end{equation}
and the value is $\approx1.7 \times 10^{-8}$ N.

\subsubsection*{A.3.1.1 Inertia force}

The inertia force is defined as:

\begin{equation}
F_{i}=-\frac{4}{3}\pi r_{p}^{3}\left(\rho_{p}\frac{du_{p}}{dt}+m_{a}\rho_{l}\left\langle \frac{Du_{p}}{dt}-\frac{Du_{f}}{Dt}\right\rangle \right)\label{eq:inertia force app}
\end{equation}

The estimated error is:

\begin{equation}
\varepsilon_{F_{i}}=\sqrt{\left(\frac{\partial F_{i}}{\partial r_{p}} \varepsilon_{r_{p}}\right)^{2}+\left(\frac{\partial F_{i}}{\partial\rho_{p}} \varepsilon_{\rho_{p}}\right)^{2}+\left(\frac{\partial F_{i}}{\partial\left\langle a_{p}-a_{f}\right\rangle }   \varepsilon_{\langle a_{p}-a_{f}\rangle }\right)^{2}+\left( \frac{\partial F_{i}}{\partial a_{p}}  \varepsilon_{a_{p}} \right)^{2}} 
\end{equation}
The value we obtain $\varepsilon_{F_{i}} \approx 1.8 \times 10^{-8}$ N.

\subsubsection*{A.3.1.1 Basset force}

The Basset force is defined as:

\begin{equation}
F_{b}=6\pi r_{p}^{2}\mu\intop_{0}^{t}\frac{d  \boldsymbol{U}_{rel}   /d\tau}{\sqrt{\pi\nu(t-\tau)}}d\tau\label{eq:basset force app}
\end{equation}
The estimated error should be calculated as follow:

\begin{equation}
\varepsilon_{F_{b}} = \sqrt{ \left( 12 \pi r_{p} \mu \int \limits_{0}^{t} \frac{d  \boldsymbol{U}_{rel}   /d\tau}{\sqrt{\pi\nu ( t-\tau) }} d\tau\varepsilon_{r_{p}} \right)^{2} + \left( \frac{6\pi r_{p}^{2}\mu}{\sqrt{ \pi \nu  (t-\tau) }}  \frac{\Delta  \boldsymbol{U}_{rel}  }{\Delta\tau} \varepsilon_{\frac{d \boldsymbol{U}_{rel}  }{d\tau}} \right)^{2} } \label{eq:basset force-1}
\end{equation}

The first term is of the order of $10^{-11}\ \mathrm{N}$ , at the
second term , $\left( \frac{6\pi r_{p}^{2}\mu}{\sqrt{\pi\nu(t-\tau)}} \frac{\Delta   \boldsymbol{U}_{rel}  }{\Delta\tau}\right)$
is of the order of $10^{-5}$ N. The value of $\left( \varepsilon_{\frac{\Delta \boldsymbol{U}_{rel} }{\Delta\tau}} \right)$
should be estimated by calculating the standard deviation of the term.
However, since the Basset force was calculated based on the previous
6 time steps and not every tracer appeared at each time step in the
trajectory of the particles, the resulting Basset force from the effect
of each tracer could not be calculated (only the relative velocity
between particle velocity and the average velocity of local tracers
could be calculated). Still, it can be assumed that the value of $\left( \varepsilon_{\frac{\Delta   \boldsymbol{U}_{rel}  }{\Delta\tau}} \right)$
is smaller than $10^{-3}$ N. Therefore, a conservative
estimation for the Basset force error can be of the order of $\sim10^{-8}$ N.
.

\end{document}